\documentclass{article}

\begin{document}

\title{Propagation of electromagnetically generated wake fields in inhomogeneous
magnetized plasmas}
\author{Martin Servin and Gert Brodin\\Department of Plasma Physics\\Ume\aa\ University\\S-901 87 Ume\aa\ , Sweden}
\maketitle
\begin{abstract}
Generation of wake fields by a short electromagnetic pulse in a plasma with an
inhomogeneous background magnetic field and density profile is considered, and
a wave equation is derived. Transmission and reflection coefficients are
calculated in a medium with sharp discontinuities. Particular attention is
focused on examples where the longitudinal part of the electromagnetic field
is amplified for the transmitted wave. Furthermore, it is noted that the wake
field can propagate out of the plasma and thereby provide information about
the electron density profile. A method for reconstructing the background
density profile from a measured wake field spectrum is proposed and a
numerical example is given.
\end{abstract}

\section{Introduction}

As is well-known, a short electromagnetic ({\small EM}) pulse propagating in
an underdense unmagnetized plasma can excite a wake field of plasma
oscillations (Tajima and Dawson 1979; Gorbunov and Kirsanov 1987). This has
interesting applications to plasma based particle accelerators (Dawson 1994),
photon acceleration (Wilks \emph{et al} 1989; Mironov \emph{et al} 1992;
Mendonca 2001) and is naturally of importance for the general understanding of
the interactions between plasmas and radiation. If the plasma is magnetized
and the external magnetic field non-parallel to the direction of propagation
of the exciting pulse, the wake field becomes partially electromagnetic and
thereby obtains a nonzero group velocity (Brodin and Lundberg 1998).

In the present paper we study wake field generation and propagation in an
inhomogeneous magnetized plasma. A wave equation for the wake field, including
arbitrary inhomogeneities in the particle number density and magnetic field,
is derived from the cold electron fluid equations and the propagation
properties are investigated. We address two questions in particular. Firstly,
we examine the effect of a strong inhomogeneity on the wake field, by
introducing a discontinuity in the background magnetic field and density. The
longitudinal part of the electric field of the transmitted wave can be largely
amplified when the ratio between the group velocities of the transmitted and
incident wave is small. The amplification factor for the longitudinal electric
field is given and analyzed as well as the transmission and reflection
coefficients. Secondly, we consider to what extent the wake field can
propagate out of the plasma, and thereby provide information about the
background plasma parameters. Since the wake field 
initially has the frequency equal to the local plasma frequency also in the 
magnetized case, this provides a way of extracting information about the 
background electron density profile, i.e. the profile in absence of the wake field 
density oscillations. It turns out that even though wave overtaking -- for example 
when a higher frequency part of the wake field passes a lower frequency part -- 
may occur, the density profile can still be reconstructed by integrating the ray equations of
geometric optics backwards. A numerical example is provided, where the
predicted spectrum of the wake field corresponding to an assumed density
profile is shown, and a reconstructed profile is calculated.

The paper is organized as follows: After stating the equations governing the
wake field in section 2, we derive the wave equation for the longitudinal
electric field in section 3. The excitation and propagation in a weakly
inhomogeneous medium are examined in section 4. Then, in section 5, the effects
of strong inhomogeneities on the wake field, which for example causes field
amplification and reflection, are studied. Next, in section 6, the spectral
properties of an electromagnetically generated wake field from a nonuniform
density profile are investigated. An algorithm for reconstructing the density
profile from a measured wake field spectrum is given and illustrated with a
numerical example. Finally, the results are summarized and discussed in
section 7.

\section{Basic equations}

We consider a high frequency {\small EM} pulse with frequency $\omega_{H}$
propagating in a cold, inhomogeneous magnetized plasma. We assume the ordering
$\omega_{H}\gg\omega_{p},\omega_{c}$, where $\omega_{p}$ and $\omega_{c}%
\equiv|q\mathbf{B}_{0}|/m$ are the plasma and electron cyclotron frequency
respectively, $q$ and $m$ are the electron charge and mass, and $\mathbf{B}%
_{0}=B_{0}\widehat{\mathbf{x}}$ is the external magnetic field. We let the
{\small EM} pulse propagate perpendicularly to the external magnetic field.
The ponderomotive force of the {\small EM} pulse will generate a ``low
frequency'' wake field mode (which is the low frequency branch of the
extraordinary mode, or plasma oscillations modified by the magnetic field,
depending on the choice of terminology) during its path through the plasma.
The generation mechanism is most efficient if the pulse has a duration of the
order of the inverse plasma frequency or shorter so that the ion-motion can be
omitted. In principle, the {\small EM} pulse will broaden due to ordinary
dispersion, decrease its energy and frequency due to the interaction with the
wake field, etc. These and other effects have been considered in homogeneous
plasmas by for example Brodin and Lundberg (1998). We will focus on the
propagation properties of the\emph{\ wake field}, however, and for this
purpose it turns out that we can forget about the details of the {\small EM}
pulse. Basically the effect of the {\small EM} field is to provide a well
localized ponderomotive source term in the governing equations for the wake
field, travelling with almost the speed of light in vacuum.

The wake field quantities are denoted by index $L$. We introduce the
corresponding vector and scalar potentials $\mathbf{A}_{L}(z,t)$ and $\phi
_{L}(z,t)$, using Coulomb gauge, and the electron density is written
$n=n_{0}(z,t)+n_{L}(z,t)$, where $n_{0}$ is the unperturbed density.
Furthermore, the electron fluid velocity $\mathbf{v}$ is divided into its
high- and low frequency part, and we denote the low frequency contributions
perpendicular and parallel to the direction of propagation with $\mathbf{v}%
_{L\perp}$ and\textbf{\ }$v_{Lz}$ respectively. The ponderomotive force of the
{\small EM} pulse induces longitudinal wake field motion, which couple to
motion in the $\widehat{\mathbf{y}}$ -direction through the Lorentz-force, but
there is no wake field motion in the direction of the external magnetic field,
and accordingly we put $\mathbf{A}_{L}=A_{L}\widehat{\mathbf{y}}$ and
$\mathbf{v}_{L\perp}=v_{L\perp}\widehat{\mathbf{y}}$. Linearizing in the low
frequency variables we obtain the following set of equations governing the
wake field generation and propagation%

\begin{eqnarray}
-\mu_{0}qn_{0}v_{L\perp}+\left[  c^{-2}\partial_{t}^{2}-\partial_{z}%
^{2}\right]  A_{L}  & = & 0\label{A_perp}\\
c^{-2}\partial_{z}\partial_{t}\phi_{L}-\mu_{0}qn_{0}\left.  v_{L}\right.
_{z}  & = & 0\label{A_par}\\
\partial_{t}\left.  v_{L}\right.  _{\perp}-\omega_{c}\left.  v_{L}\right.
_{z}+\frac{q}{m}\partial_{t}A_{L}  & = & 0\label{mom_perp}\\
\partial_{t}\left.  v_{L}\right.  _{z}+\frac{q}{m}\partial_{z}\phi_{L}%
+\omega_{c}\left.  v_{L}\right.  _{\perp} & = & -\frac{q^{2}}{2m^{2}}%
\partial_{z}|\mathbf{A}_{H}|^{2}\label{mom_par}\\
\partial_{t}n_{L}+\partial_{z}(n_{0}\left.  v_{L}\right.  _{z})  & = & 0\label{cont}%
\end{eqnarray}

\section{Derivation of the wave equation}

Before deriving a wave equation for the wake field it is practical to redefine
$\frac{q}{m}A_{L}\rightarrow A$, $\frac{q}{m}\partial_{z}\phi\rightarrow\psi$,
$v_{\perp}\rightarrow v$ and $S=-(q^{2}/2m^{2})\partial_{z}|\mathbf{A}%
_{H}|^{2}$. Eqs. (\ref{A_perp}), (\ref{mom_perp}) and (\ref{mom_par}) then reads%

\begin{eqnarray}
-\omega_{p}^{2}v+\left[  \partial_{t}^{2}-c^{2}\partial_{z}^{2}\right]  A  &
= & 0\label{nA_perp}\\
\partial_{t}v-\omega_{c}\omega_{p}^{-2}\partial_{t}\psi+\partial_{t}A  &
= & 0\label{nmom_perp}\\
\left[  \omega_{p}^{-2}\partial_{t}^{2}+1\right]  \psi+\omega_{c}v  &
= & S\label{nmom_par}%
\end{eqnarray}
Eq. (\ref{A_par}) and (\ref{cont}) only give information of how $\left.
v_{L}\right.  _{z}$ and $n_{L}$ are related to the other variables and are
therefore omitted at this point. Acting on Eq. (\ref{nmom_perp}) with $\left[
\partial_{t}^{2}-c^{2}\partial_{z}^{2}\right]  \partial_{t}^{-1}$ and applying
Eq. (\ref{nA_perp}) gives
\[
0=\left[  \partial_{t}^{2}-c^{2}\partial_{z}^{2}+\omega_{p}^{2}\right]
v-\left[  \partial_{t}^{2}-c^{2}\partial_{z}^{2}\right]  (\omega_{c}\psi)
\]
Combining this with Eq. (\ref{nmom_par}) gives after some rearrangements
\begin{equation}
\partial_{z}^{2}(\hat{\alpha}\psi)+\hat{\beta}\psi=\hat{\gamma}(\omega
_{c}^{-1}S)\label{wave_eq}%
\end{equation}
where
\begin{eqnarray*}
\hat{\alpha}  & \equiv & c^{2}\left[  \omega_{c}^{-1}(1+\omega_{p}^{-2}%
\partial_{t}^{2})+\omega_{c}\omega_{p}^{-2}\right] \\
\hat{\beta}  & \equiv & -\omega_{c}^{-1}\left[  \omega_{p}^{2}+\partial_{t}%
^{2}\left(  2+\omega_{p}^{-2}(\omega_{c}^{2}+\partial_{t}^{2})\right)  \right]
\\
\hat{\gamma}  & \equiv & -[\partial_{t}^{2}-c^{2}\partial_{z}^{2}+\omega_{p}^{2}]
\end{eqnarray*}

Eq. (\ref{wave_eq}) provides the starting point for analyzing the propagation
properties of a wake field generated by an {\small EM} pulse under the given
circumstances. It is somewhat surprising that the evolution of the wake field
is governed by a \emph{single} wave equation, in spite that there are two
arbitrary background parameter functions $\omega_{p}$ and $\omega_{c}$.

In the case of a static and homogeneous background density and magnetic field
this reduces to
\begin{equation}
\left[  \partial_{t}^{4}+(\omega_{p}^{2}+\omega_{h}^{2})\partial_{t}^{2}%
-c^{2}\partial_{z}^{2}\partial_{t}^{2}-\omega_{h}^{2}c^{2}\partial_{z}%
^{2}+\omega_{p}^{4}\right]  \psi=\omega_{p}^{2}[\partial_{t}^{2}-c^{2}%
\partial_{z}^{2}+\omega_{p}^{2}]S\label{hom_eq}%
\end{equation}
where $\omega_{h}^{2}\equiv\omega_{p}^{2}+\omega_{c}^{2}$ is the upper hybrid
frequency. The left-hand side of Eq. (\ref{hom_eq}) is the familiar wave
operator for the extraordinary electromagnetic mode.

\section{Wake field excitation and propagation}

We will consider wake field propagation in both strongly and weakly
inhomogeneous plasmas. By \emph{weakly} inhomogeneous we mean that the
wavelength of the wake field is much smaller than the characteristic
inhomogeneity length scale. Then, to lowest order, derivatives acting on
background quantities can be neglected and the wave equation reduces to Eq.
(\ref{hom_eq}) with a space and possibly a time dependence in $\omega_{p}$ and
$\omega_{c}$. In a \emph{strongly} inhomogeneous plasma this approximation
cannot be applied, and the time evolution of the wake field is given by Eq.
(\ref{wave_eq}). In this section only weakly inhomogeneous plasmas will be
considered, and it is illustrative to divide our study of the wake field
properties into its excitation and its propagation phase. However, we note
that the excitation process considered below will also be of relevance for the
next section concerning strongly inhomogeneous plasmas, since in that case we
will deal with wake fields {\it generated} in a weakly inhomogeneous plasma that
{\it propagates} into a strongly inhomogeneous region.

\subsection{Excitation}

The excitation of one additional wavelength of the wake field takes place
during a distance of the order of $2\pi c/\omega_{p}$, and -- as a basic
assumption of ours -- the variations of $n_{0}$ is negligible on this length
scale. Thus as far as the excitation process is concerned, the plasma can
essentially be treated as homogeneous. The solution for the wake field can
thus be obtained from previous authors (Brodin and Lundberg 1998). Changing to
co-moving coordinates $\xi=z-v_{gH}t$, $\tau=t$, where $v_{gH}$ is the group
velocity of the high frequency {\small EM} pulse, in Eq. (\ref{hom_eq}) and
neglecting the small derivatives $\partial_{\tau}^{2}$ and $\partial_{\tau
}\partial_{\xi}$ and terms proportional to $v_{gH}^{2}-c^{2}$ it reduces to
(reinstating the potentials)
\begin{equation}
\left[  v_{g{\small H}}^{2}\partial_{\xi}^{2}+\omega_{p}^{2}\right]  \phi
_{L}=\frac{q\omega_{p}^{2}|\mathbf{A}_{H}|^{2}}{m}%
\end{equation}
This implies
\begin{equation}
\phi_{L}=\phi_{L0}\sin[k_{p}(\xi-\xi_{0})]\label{initial}%
\end{equation}
where $k_{p}$ is the wake field wavenumber $k_{p}=\omega_{p}/v_{gH}$, $\xi
_{0}$ is the (constant) position of the (short) {\small EM} pulse, and
$\phi_{L0}=(q\omega_{p}/mv_{gH})\int_{-\infty}^{\infty}|\mathbf{A}_{H}%
|^{2}d\xi$. The important result here, for our purposes, is the determination
of the initial value of the wake field wave number $k_{p}=\omega_{p}/v_{gH}$,
which corresponds to an initial frequency $\omega_{p}$ (in the laboratory
frame). Note that this wake field frequency will in general vary with the
position of generation.

\subsection{Propagation}

In a weakly inhomogeneous plasma the wavelength of the wake field is much
smaller than the characteristic inhomogeneity length scale. The wake field
properties can thus be considered as locally uniform but globally nonuniform,
and therefore we make the ansatz of geometrical optics (Whitham 1974)
\[
\psi=\psi_{0}(z,t)e^{i\theta(z,t)}%
\]
The local wavenumber and frequency are defined in terms of the eikonal
$\theta(z,t)$ as $k\equiv\partial_{z}\theta$ and $\omega\equiv-\partial
_{t}\theta$, respectively and the amplitude $\psi_{0}(z,t)$ is assumed to vary
slowly with $z$ and $t$. The local dispersion relation follows then, as a
lowest order approximation, from Eq. (\ref{hom_eq})
\begin{equation}
\omega^{4}-\omega^{2}(\omega_{h}^{2}+\omega_{p}^{2}+c^{2}k^{2})+\omega_{h}%
^{2}c^{2}k^{2}+\omega_{p}^{4}=0\label{disprel}%
\end{equation}
We note that there is a resonance at $\omega^{2}=\omega_{h}^{2}\equiv
$\ $\omega_{c}^{2}+\omega_{p}^{2}$\ and cut-offs at $\omega_{L}\equiv$
$\frac{1}{2}[-\omega_{c}+(\omega_{c}^{2}+4\omega_{p}^{2})^{1/2}]$ and
$\omega_{R}\equiv$ $\frac{1}{2}[\omega_{c}+(\omega_{c}^{2}+4\omega_{p}%
^{2})^{1/2}]$. The dispersion relation has two positive roots. One branch is
valid for $\omega>\omega_{R}$ and one for $\omega_{L}<\omega<\omega_{h}$. As
the wake field is generated with the local plasma frequency $\omega_{p},$ the
wake field must belong to the latter branch. Therefore we write the dispersion
relation from now on as
\begin{equation}
\omega=W(k,z,t)\equiv\left\{  \chi-\sqrt{\chi^{2}-\omega_{h}^{2}c^{2}%
k^{2}-\omega_{p}^{4}}\right\}  ^{1/2}\label{omega}%
\end{equation}
where $\chi\equiv(\omega_{h}^{2}+\omega_{p}^{2}+c^{2}k^{2})/2$. From the
dispersion relation we also derive explicit expressions for the wavenumber and
group velocity $v_{g}\equiv\partial_{k}\omega$ that we will use later. They
are
\begin{equation}
k=\frac{1}{c}\left[  \frac{\omega^{2}(\omega_{h}^{2}+\omega_{p}^{2}%
)-\omega^{4}-\omega_{p}^{4}}{\omega_{h}^{2}-\omega^{2}}\right]  ^{1/2}%
\label{k}%
\end{equation}
and
\begin{equation}
v_{g}=\frac{\omega^{2}-\omega_{h}^{2}}{2\omega^{2}-\omega_{h}^{2}-\omega
_{p}^{2}-c^{2}k^{2}}\frac{c^{2}k}{\omega}\label{vg}%
\end{equation}
respectively.

From the geometric optics approach there follows useful transport equations
for the wavenumber and frequency. Noting the identity $\partial_{t}%
k+\partial_{z}\omega=0$ it follows that
\begin{equation}
\frac{dk}{dt}=-\partial_{z}W\quad,\quad\quad\frac{d\omega}{dt}%
=\partial_{t}W\label{rayeqs}%
\end{equation}
where $d/dt\equiv\partial_{t}+v_{g}\partial_{z}$. Eqs. (\ref{rayeqs}) are
referred to as the \emph{ray equations}. Note that for a time-independent
medium, the right hand side of the last equation is zero, and the wake field
propagates with the local wake field group velocity $v_{g}$ and with
unchanged frequency.

Since the group velocity may vary along the rays, the energy carried with the
wake field can be compressed as well as attenuated, and from energy
conservation one may expect the field amplitude to vary correspondingly, see
e.g. Mendonca (2001) for similar effects for ordinary electromagnetic waves in
plasmas. The extraordinary mode, however, has several degrees of freedom that
the energy may vary between, depending on variations in the background
parameters. Therefore the behavior of the amplitudes are not in direct
correspondence with the variation of the group velocity. For completeness and
for future reference we state the linear relations between the field variables
and $\psi$ that follows from Eqs. (\ref{A_perp})-(\ref{cont}) in the weakly
inhomogeneous approximation
\begin{eqnarray}
\phi & = & -ik^{-1}\frac{m}{q}\psi\\
v_{\perp} & = & \frac{\omega_{p}^{-2}\omega_{c}(\omega^{2}-c^{2}k^{2})}%
{\omega^{2}-c^{2}k^{2}-\omega_{p}^{2}}\psi\label{lin_v_perp}\\
v_{z} & = & -i\omega\omega_{p}^{-2}\psi\\
A_{L} & = & -\frac{m}{q}\frac{\omega_{c}}{\omega^{2}-c^{2}k^{2}-\omega_{p}^{2}%
}\psi\\
\delta n & = & -ikn\omega_{p}^{-2}\psi
\end{eqnarray}

\section{Reflection and transmission properties}

We now consider the effect of a strongly inhomogeneous region on the wake
field. We assume that the wake field entering this region was generated in a
weakly inhomogeneous part of the plasma. Thus the wake field can be taken to
be uniform when entering the inhomogeneous region, and the variations in the
wake field frequency $\omega$ can be neglected. During these conditions we
have $\partial_{t}^{2}\psi=-\omega^{2}\psi$ and, away from the exciting
electromagnetic pulse, the wave equation (\ref{wave_eq}) reduces to an
ordinary differential equation
\begin{equation}
\partial_{z}^{2}(\alpha\psi)+\beta\psi=0\label{ode}%
\end{equation}
where
\begin{eqnarray*}
\alpha & \equiv & c^{2}\omega_{c}^{-1}\omega_{p}^{-2}(\omega_{h}^{2}-\omega
^{2})\\
\beta & \equiv & -\omega_{c}^{-1}\omega_{p}^{-2}(\omega^{4}+\omega_{p}^{4}%
-\omega^{2}(\omega_{h}^{2}+\omega_{p}^{2}))
\end{eqnarray*}

Although it is straight forward, at least numerically, to solve Eq.
(\ref{ode}) for any given background density and magnetic field, we simplify
the analysis by treating the inhomogeneity as a discontinuity, in order to
clearly illustrate some of the main effects associated with a strong
inhomogeneity. We let the discontinuity be located at $z=0$, and the remaining
plasma is assumed to be homogeneous. Thus we can make the following ansatz for
the wake field
\begin{equation}%
\begin{array}
[c]{lll}%
\psi_{1} & =\psi_{i}e^{i(k_{1}z-\omega t)}+\psi_{r}e^{i(-k_{1}z-\omega t)} &
\quad,\quad\quad z<0\\
\psi_{2} & =\psi_{t}e^{i(k_{2}z-\omega t)} & \quad,\quad\quad z>0
\end{array}
\label{ansatz}%
\end{equation}
where $\omega=\omega_{p1}$. This ansatz does not apply if we are too close to
the exciting pulse, or for the fields that were generated in the strongly
inhomogeneous region, but both these parts of the wake field are assumed to be
distant to the discontinuity. The subscripts $i$, $r$ and $t$ stands for the
incident, reflected and transmitted part, respectively, and the indices $1$
and $2$ distinguishes quantities on the left ($z<0$) and right ($z>0$) hand
side of the discontinuity.

By integrating Eq. (\ref{ode}) across $z=0$, it follows that $\alpha\psi$ and
$\partial_{z}(\alpha\psi)$ are continuous over the discontinuity. We define
$r=\psi_{r}/\psi_{i}$ and $a=\psi_{t}/\psi_{i}$, and we refer to these
quantities as the amplification factors for the reflected and transmitted part
of the longitudinal electric field, which essentially are generalized Fresnel
coefficients. The continuity conditions and the ansatz (\ref{ansatz}) imply
\begin{equation}
a=\frac{2k_{1}}{k_{1}+k_{2}}\frac{\alpha_{1}}{\alpha_{2}}\label{a}%
\end{equation}
where $\alpha_{1}$ and $\alpha_{2}$ are the values of $\alpha$ on the left and
right hand side of the discontinuity, respectively, and
\begin{equation}
r=\frac{k_{1}-k_{2}}{k_{1}+k_{2}}\label{r}%
\end{equation}
It follows from Eq. (\ref{ode}) that
\begin{equation}
S=-\frac{\omega k\omega_{c}^{2}}{\left(  \omega^{2}-c^{2}k^{2}-\omega_{p}%
^{2}\right)  ^{2}}|\psi|^{2}\label{Poynting}%
\end{equation}
is -- averaged in time and space -- a conserved quantity, i.e. $\partial
_{z}S=0$. (Noting that $S$ is the time and space averaged z-component of the
Poynting vector $\mathbf{S}\equiv\frac{1}{2}[\mathbf{E}\times\mathbf{B}^{\ast
}+\mathbf{E}^{\ast}\times\mathbf{B}]$, this also follows directly from energy
conservation. Actually from Eq. (\ref{ode}) it firstly follows that
the conserved quantity is equal to $k\alpha^{2}|\psi|^{2}$. To see that this is equivalent 
to Eq. (\ref{Poynting}) requires tedious but straight forward algebra.)

The transmission and reflection coefficients are introduced as $T\equiv
S_{t}/S_{i}$ and $R\equiv S_{r}/S_{i}$. Explicitly they read
\begin{equation}
T=\frac{4k_{1}k_{2}}{\left(k_{1}+k_{2}\right)  ^{2}}
\end{equation}
and
\begin{equation}
R=\frac{\left(  k_{1}-k_{2}\right)  ^{2}}{\left(  k_{1}+k_{2}\right)  ^{2}%
}\label{R}%
\end{equation}
and they satisfy the energy conservation law $T+R=1$.

The quantities $a$, $r$, $T$ and $R$ depends on the four parameters
$\omega_{p1}$, $\omega_{p2}$, $\omega_{c1}$ and $\omega_{c2}$. We choose
$\omega_{p1}=$ $\omega_{c2}=1$ (in normalized units) and, rather than
presenting complicated surface plots, present one dimensional graphs of the
dependence of $a$, $r$, $T$ and $R$ on $\omega_{c2}$ for some given values of
$\omega_{p2}$, and vice versa. Fig. 1 shows $T(\omega_{c2})$, $R(\omega_{c2})$
and $v_{g2}$ for distinct values of $\omega_{p2}$, and Fig. 2 shows
$T(\omega_{p2})$, $R(\omega_{p2})$ and $v_{g2}$ for distinct values of
$\omega_{c2}$. As can be seen, for most of the parameter regime the
transmission is close to unity. There are also regions of no propagation in
Fig. 2, corresponding to parameter values $\omega_{h2}<\omega<\omega_{R2}$.
Close to these regions the transmission quickly goes to zero and the
reflection towards unity, because the wave approaches a resonance or a
cut-off. The regions to the left in Fig. 1a and 2a corresponds to the branch
$\omega>\omega_{R2}$, i.e. the transmitted mode belongs to a different branch
of the dispersion relation (\ref{disprel}) than the incident mode.

In Fig. 1a, where $\omega_{p2}=0.5$, the left part stretches until
$\omega_{c2}\approx0.71$ where the $\omega_{R}$ cut-off prohibits
transmission. At $\omega_{c2}\approx0.87$, $T$ and $v_{g2}$ becomes zero
because of the $\omega_{h}$ resonance. Fig. 1b is the special case for a jump
in the magnetic field only, i.e. $\omega_{p2}=1$. The transmission is
everywhere unity although the group velocity approaches zero with diminishing
$\omega_{c2}$. In Fig. 1c, where $\omega_{p2}=2$, there is no transmission up
to $\omega_{c2}=3$ which is due to the $\omega_{L}$ cut-off.

In Fig. 2a, 2b and 2c the values of $\omega_{c2}$ are $0.5$, $1$ and $2$,
respectively. In Fig. 2a the $\omega_{R}$ cut-off occurs at $\omega
_{p2}\approx0.71$, the $\omega_{h}$ resonance at $\omega_{p2}\approx0.87$ and
the $\omega_{L}$ cut-off at $\omega_{p2}\approx1.22$. In both Fig 2b and 2c
the absence of transmission is due to the $\omega_{L}$ cut-off.

The special cases Fig. 1b and 2b are particularly interesting, representing a
jump in the magnetic field only (referred to as case I below) and in the
density only (referred to as case II below). The transmission remains unity
although the group velocity goes to zero as $\omega_{c2}\rightarrow0$ (Fig.
1b) and $\omega_{p2}\rightarrow0$ (Fig. 2b). This means that the energy
density entering from region 1 will be dramatically amplified in region 2. The
question is to which field variables this energy will be concentrated.

\emph{Case I.} For $\omega_{p2}=\omega_{p1}=1$ the amplification factor for
$\psi$ is displayed in Fig. 3a together with $v_{g}$ and $v_{ph}$. The
amplification factor reduces to $a=\omega_{c1}/\omega_{c2}$. It should be
emphasized that the wavenumber is preserved over the discontinuity and thus
also the phase velocity, that equals $c$, is preserved. This property is,
however, very sensitive to small deviations from exactly constant density. This
is illustrated in Fig. 3b, where $\omega_{p2}=0.99$, and 3c, where
$\omega_{p2}=1.01$. In these cases the longitudinal field can still be
amplified, but not by a large factor without also affecting the phase velocity substantially.

\emph{Case II.} In the case of uniform magnetic field, $\omega_{c2}%
=\omega_{c1}=1$, the field variable that is amplified is the perpendicular
electron fluid velocity $v_{\perp}$. This can be seen by evaluating
$v_{\perp2}/v_{\perp1}$ using Eq. (\ref{lin_v_perp}) and the amplification
factor (\ref{a}). This case does not share the property with case I of
preserved phase velocity over the discontinuity. In Fig. 2b we have added a
small deviation to $\omega_{c}$ so that $\omega_{c2}=1.005$ to illustrate that
the group velocity can be made arbitrarily small. But, in the limit
$\omega_{p2}\rightarrow0$ the group velocity approaches $c$, as required in vacuum.

\section{Density profile reconstruction}

A wake field generated by a short {\small EM} pulse in an underdense
magnetized plasma has the frequency equal to the local plasma frequency,
$\omega_{p}=\sqrt{q^{2}n_{0}/\epsilon_{0}m}$. Due to the presence of the
magnetic field it has a nonzero group velocity, and for suitable background
parameter profiles cut-offs and resonances in the plasma are avoided and thus
the wake field can propagate out of the plasma. This suggests the possibility
of gaining information of the density profile $n_{0}(z)$ from studying the
wake field exiting the plasma.

We assume that the plasma is weakly inhomogeneous so that the results in
section 4.1 and 4.2 can be applied. Given the ray equations, one may --
conceptually speaking -- treat the wake field as consisting of particles,
``photons'' (dressed photons, that is), created at different times, as the
{\small EM} pulse plows through the plasma. The ``photons'' have distinct
frequencies, depending on the position they were ``created'', and thus also
different propagation velocity, $v_{g}$. The equation of motion for the
``photons'' are Eq. (\ref{vg}) and the ray equations, Eq. (\ref{rayeqs}). The
exact evolution of the wake field, on the other hand, is more complicated.
This can be understood from that ``photons'' created at later times may be
overtaken by ``photons'' created at earlier times. For our purposes the
particle-picture is to prefer.

We assume that the generated wake field spectrum is measured immediately
outside the plasma boundary. In consistence with the geometric optics
approximation we will treat the measured data as a weakly time dependent
spectrum with well defined sharp (quasi-monochromatic) peaks. Because of
overtaking ``photons'', the data is not necessarily monochromatic at a given
time, multiple sharp peaks may occur in the spectrum. Generally, we can
express the data as a set of distinct frequencies measured at different times,
in which case a sharp curve can be recognized, see Fig. 4. Due to cut-offs
and/or resonances, the curve may be discontinuous.

Given a measured wake field spectrum, as in Fig. 4, the density profile can be
reconstructed the following way. Discretize the frequency curve into $N$
points $\omega^{(0)},\omega^{(1)},...,\omega^{(N)}$ with corresponding time of
detection $t^{(0)},t^{(1)},...,t^{(N)}$. The plasma is discretized into $N$
cells whose positions, $z^{(0)},z^{(1)},...,z^{(N)}$ and width are yet to be
determined. The ``photon'' with frequency $\omega^{(N)}$, detected at time
$t^{(N)}$, was the last one to exit the plasma. Therefore the plasma frequency
in the cell at $z^{(N)}$ has the value $\omega^{(N)}$. Next, we retrace the
``photon'' $\omega^{(N-1)}$ backwards into the plasma to a position
$z^{(N-1)}$ consistent with the time of detection $t^{(N-1)}$ and the equation
of motions through the already reconstructed cell. The plasma frequency in the
cell with position $z^{(N-1)}$ is assigned the value $\omega^{(N-1)}$. The
third ``photon'' $\omega^{(N-2)}$ is retraced through the cells with plasma
frequency $\omega^{(N)}$ and $\omega^{(N-1)}$ to a position $z^{(N-2)}$ and
assigns the corresponding cell there the plasma frequency $\omega^{(N-2)}$.
This procedure is repeated for all ``photons'' along the frequency curve.

In order to demonstrate the method we numerically calculate a wake field
spectrum from an assumed density profile, using the ray equations, see Fig.5(a), and treat this as
experimental data from which the density profile can be reconstructed. We
consider a plasma magnetized such that $\omega_{c}=1.1\times\omega
_{p,\textrm{max}}$, where $\omega_{p,\textrm{max}}$ is the maximum value of the
plasma frequency. For simplicity we normalize such that $L=1$, $c=1$, where
$L$ is the length of the plasma, and let the {\small EM} pulse enter the
plasma at $t=0$ and exit at $t=1$.

Retracing the spectrum according to the algorithm presented above results in a
density profile that can be compared with the one we assumed, see Fig. 5(b). A
small numerical error -- that can be removed with a finer discretization --
can be seen. Note that the entire plasma profile cannot be reconstructed. The
left most points in Fig. 2(b) are missing. This is because the wake field
generated in this region of low density cannot propagate through the plasma
since there is a cut-off prohibiting this. The information of this region is
already missing in the wake field spectrum.

\section{Summary and discussion}

We have considered the propagation of wake fields generated by a short, high
frequency {\small EM} pulse in an inhomogeneous magnetized plasma. A general
wave equation for the wake field driven by the ponderomotive force of the high
frequency pulse has been derived, Eq. (\ref{wave_eq}), and the propagation
properties have been investigated. If the wake field enters a strongly
inhomogeneous region it may be largely amplified. The amplification factor
for the longitudinal electric field and transmission and reflection
coefficients have been derived and analyzed for a discontinuity in the
magnetic field and/or the particle number density. In the case of uniform
density the amplification factor becomes $a=\omega_{c1}/\omega_{c2} $. This
special case is also characterized by the fact that there is no reflection and
the phase velocity of the transmitted wave remains equal to the velocity of
light. This result may be of relevance for particle and photon accelerators
based on wake fields. For applications like particle acceleration, it might be
desirable to finally let the amplified wake field propagate in an
unmagnetized plasma. Simply eliminating the external magnetic field once the
wake field is amplified affects the frequency and thus the phase velocity in
accordance with the ray equations (\ref{rayeqs}). It seems straightforward,
however, to match the discontinuity and the elimination of the magnetic field
so that the phase velocity of the resulting field is approximately equal to
$c$, which is the desirable value for particle acceleration purposes.

Furthermore, the spectral properties of a wake field from a plasma with
nonuniform density have been investigated and a method for reconstructing the
density profile from a measured wake field spectrum has been proposed and
illustrated with a numerical example. This result shows that wake fields
generated by a high frequency {\small EM} pulse in principle can be used as a
diagnostic tool, in magnetized plasmas. The proposed method is based on a
mechanism substantially different from those of existing techniques, such as
interferometry and reflectometry (Hartfuss 1998; Hutchinson 1987). It should
be possible to extend the method to plasmas where the background density
varies in time by using sequential {\small EM} pulses. The most interesting
case is that of a strongly magnetized plasma (i.e. when the electron cyclotron
frequency is larger than the plasma frequency), for which almost all of the
wake field energy - except that generated in a narrow low density region - may
propagate out of the plasma. The requirement that the length scales of
inhomogeneities must be larger than the local plasma wave length $\lambda
_{p}\equiv2\pi v_{gH}/\omega_{p}$ for the results to be valid means that it
can resolve inhomogeneities of the order $10^{8}n^{-1/2}\mathrm{m}$ and
larger, where $n$ is the electron number density in units \textrm{m}$^{-3}$.

\section{References}

G. Brodin and J. Lundberg 1998 \emph{Phys. Rev. E} \textbf{57},
704.\newline J. M. Dawson 1994 \emph{Phys. Scr.} \textbf{T 52}, 7.\newline L.
M. Gorbunov and V. I. Kirsanov 1987 \emph{Zh.Eksp. Theor. Fiz.} \textbf{93},
509. [1987 \emph{Sov.Phys. JETP }\textbf{66}, 290].\newline H. J. Hartfuss
1998 \emph{Plasma Phys. Control. Fusion }\textbf{40} A231-A250.\newline I. H.
Hutchinson 1987 \emph{Principles of plasma diagnostics} Cambridge University
Press, Cambridge.\newline J. T. Mendonca 2001 \emph{Theory of Photon
Acceleration} Institute of Physics Publishing, Bristol.\newline V. A. Mironov,
A. M. Sergeev, E. V. Vanin, G. Brodin and J. Lundberg 1992 \emph{Phys. Rev. A}
\textbf{46}, 6178.\newline T. Tajima and J. M. Dawson 1979 \emph{Phys. Rev.
Lett. A} \textbf{43}, 276.\newline G. B Whitham 1974 \emph{Linear and
Nonlinear Waves }John Wiley \& Sons, New York.\newline S. C. Wilks, J. M.
Dawson, W. B. Mori, T. Katsouleas and M. E. Jones 1989 \emph{Phys. Rev. Lett.}
\textbf{62}, 2600.

\newpage

\section{Figure Captions}

\textbf{Figure 1.} Reflectivity, transmittivity and transmitted group velocity
as a function of the jump in external magnetic field for $\omega_{p2}%
/\omega_{p1}=0.5$. (Fig. 1a), for $\omega_{p2}/\omega_{p1}=1$ (Fig. 1b) and
for $\omega_{p2}/\omega_{p1}=2$ (Fig. 1c), while $\omega_{c1}/\omega_{p1}=1$
in all three figures.

\bigskip

\textbf{Figure 2.} Reflectivity, transmittivity and transmitted group velocity
as a function of the jump in the unperturbed density for $\omega_{c2}%
/\omega_{c1}=0.5$. (Fig. 2a), for $\omega_{c2}/\omega_{c1}=1$ (Fig. 2b) and
for $\omega_{c2}/\omega_{c1}=2$ (Fig. 2c), while $\omega_{c1}/\omega_{p1}=1$
in all three figures.

\bigskip

\textbf{Figure 3.} Longitudinal field amplification, phase velocity and group
velocity of the transmitted wave as a function of the jump in external
magnetic field for $\omega_{p2}/\omega_{p1}=0.99$. (Fig. 3a), for $\omega
_{p2}/\omega_{p1}=1$ (Fig. 3b) and for $\omega_{p2}/\omega_{p1}=1.01$ (Fig.
3c), while $\omega_{c1}/\omega_{p1}=1$ in all three figures.\bigskip

\textbf{Figure 4.} Cartoon picture of a detected wake field spectrum as a
function of time of detection. The figure also illustrates the disctretization
of the curve that can be identified from a wake field spectrum.\bigskip

\textbf{Figure 5.} Example of a numerically generated wake field spectrum
(Fig. 5a) from an assumed density profile (solid line in Fig. 5b). The
reconstructed profile is marked with crosses (Fig. 5b).
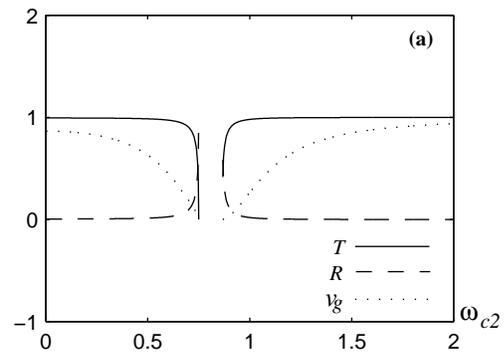
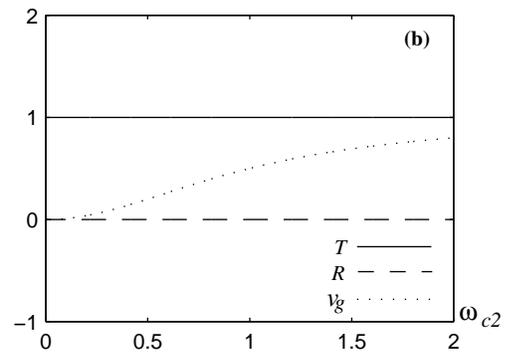
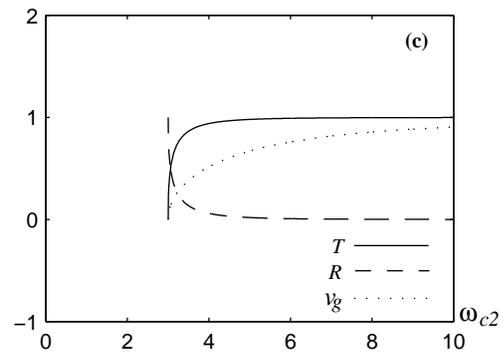
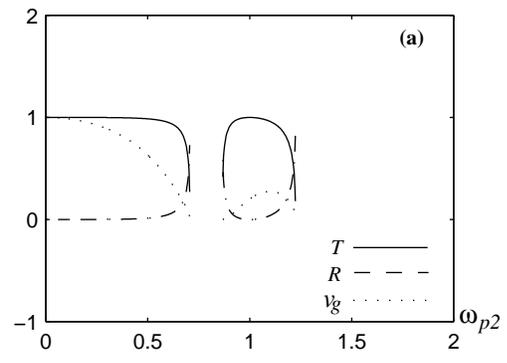
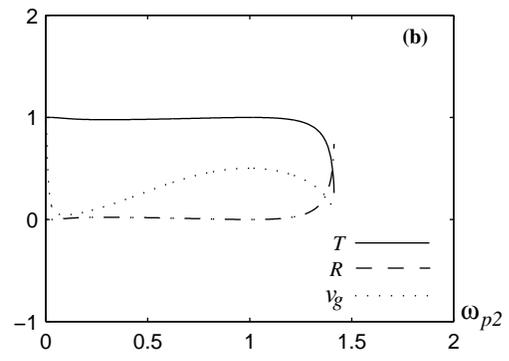
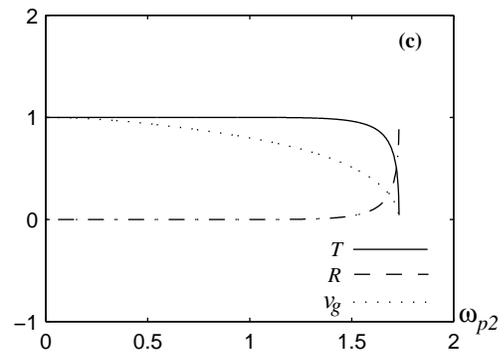
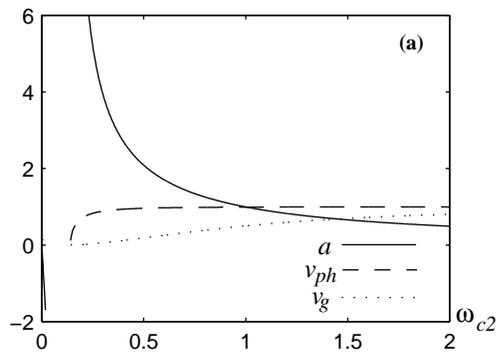
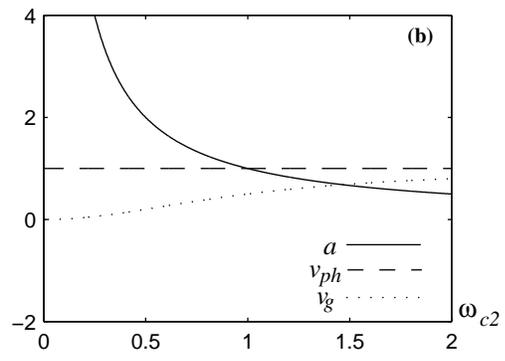
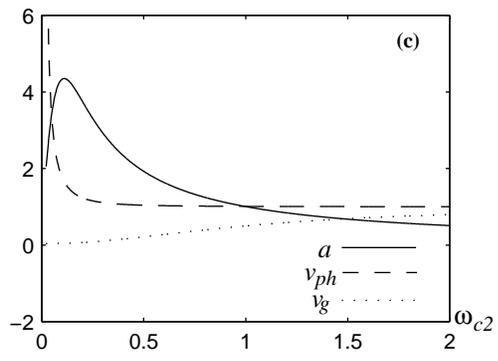
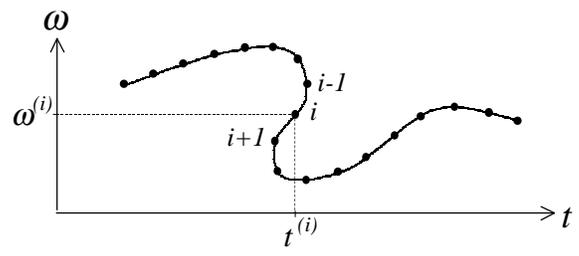
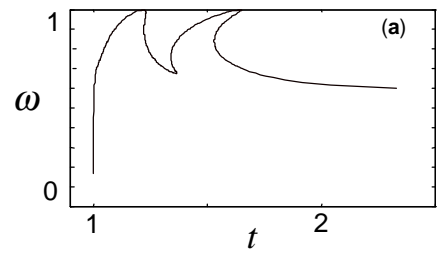
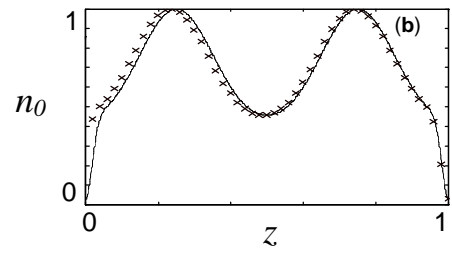
\end{document}